\documentclass[prb,twocolumn]{revtex4}

\usepackage{amssymb}
\usepackage{amsmath}
\usepackage{graphicx}
\usepackage{amsthm,amssymb}
\usepackage[mathscr]{eucal}

\def\bea{\begin{eqnarray}}
\def\eea{\end{eqnarray}}
\def\ri{{\rm i}}
\def\re{{\rm e}}
\def\rd{{\rm d}}

\def\Xint#1{\mathchoice
    {\XXint\displaystyle\textstyle{#1}}%
    {\XXint\textstyle\scriptstyle{#1}}%
    {\XXint\scriptstyle\scriptscriptstyle{#1}}%
    {\XXint\scriptscriptstyle\scriptscriptstyle{#1}}%
    \!\int}
\def\XXint#1#2#3{{\setbox0=\hbox{$#1{#2#3}{\int}$}
    \vcenter{\hbox{$#2#3$}}\kern-.5\wd0}}

\def\dashint{\Xint-}

\begin{document}

\begin{flushleft}
RUNHETC-06-02
\end{flushleft}

\title{Universal scaling behavior of the single electron box in the strong tunneling limit}
\author{Sergei L. Lukyanov}
\affiliation{NHETC, Department of Physics and Astronomy,
     Rutgers University,
     Piscataway, NJ 08855-0849, USA\\
and\\
L.D. Landau Institute for Theoretical Physics
  Chernogolovka, 142432, Russia}
\author{Philipp Werner}
\affiliation{Department of Physics, 
Columbia University, 538 West, 120th Street, New York, NY 10027}
\date{June 16, 2006}

\begin{abstract}

We perform a numerical analysis of recently 
proposed scaling functions for the single electron box. 
Specifically, we study the 
``magnetic'' susceptibility as a function of tunneling conductance
and gate charge, and 
the effective charging energy at zero gate charge as a function of tunneling conductance
in the strong tunneling limit.
Our Monte Carlo results 
confirm the accuracy of the theoretical predictions.

\end{abstract}

\maketitle

\section{Introduction}

\begin{figure}[ht]
\centering
\includegraphics [angle=0, width= 5cm] {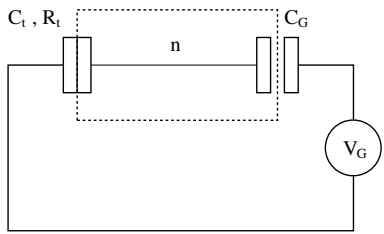}
\caption{Circuit diagram
of the single electron box.
The box with excess charge $n$ is indicated by
the dashed rectangle.
It is connected to a voltage
source through a capacitor $C_G$
and a tunnel junction with resistance $R_t$ and capacitance $C_t$.}
\label{box}
\end{figure}

A single electron box is
a low-capacitance metallic island, connected to an outside lead
by a tunnel junction. It has first been experimentally 
realized by Lafarge and co-workers [\onlinecite{Lafarge}]. 
Such a device exhibits 
Coulomb-blockade phenomena due to 
the large charging energy, which 
influences single electron tunneling.
By applying a gate voltage $V_G$ it is possible to 
induce a continuous polarization charge $n_G=C_GV_G/e$ on 
the island (see Fig.\,\ref{box}). For a very small 
dimensionless tunneling conductance  $\alpha=h/( 2\pi^2 e^2 R_t)\ll 1 $
($R_t$ is the resistance of the contact)
the energy spectrum  is given by the set of parabolas 
$E_n(n_G) = E_C(n-n_G)^2$, where the 
integer $n$ denotes the number of excess electrons. 
$E_C=e^2/2(C_t+C_G)$ is the single electron  charging energy, 
which sets the energy scale.
As the tunneling conductance is increased, quantum 
fluctuations lead to 
a renormalized {\it effective} charging energy $E_C^*$. 
The properties of the electron box for $\alpha\lesssim 0.8$
are well described 
by perturbation theory in $\alpha$, and these 
results have been verified numerically [\onlinecite{Goeppert}]. 

Over the last decade, the  analytical 
expressions describing a single electron box 
in the limit of large tunneling conductance have 
been the subject of controversial debate 
[\onlinecite{Panyukov&Zaikin, Wang&Grabert, Hofstetter, Larkin,Lukyanov}]. 
Numerical investigations which were performed 
to test the numerous incompatible predictions have proven 
difficult or inconclusive, and in some cases even added to the 
confusion [\onlinecite{Hofstetter, Wang&Egger, Koenig, Herrero, Werner05}]. 
It is the purpose of this paper to remedy this unsatisfactory state 
of affairs and present a comprehensive study of recently 
obtained scaling formulas, using state-of-the-art numerical techniques.

\section{Theory}

We address the  limit
of  large   tunneling conductance
in the framework of the Ambegaokar-Eckern-Sch\"on  model [\onlinecite{AES,Schoen}].
The partition function for the
model can be written as a
path integral over an  angular
variable $\phi$
(conjugate to the number of excess charges on the island) 
\begin{equation}
Z=\int{\cal D}\phi\  \re^{-{\mathscr   A}[\phi]}\ ,
\label{partitionfunction}
\end{equation}
where the Matsubara effective action reads 
($\hbar=1$ and $\beta$ denotes the inverse temperature)
\bea
{\mathscr   A}[\phi]&=&
{ \frac{1}{4E_C}}\int_0^\beta \rd\tau \,  (\partial_{\tau}\phi)^2+\ri\, n_G\,
\int_0^\beta \rd\tau \,\partial_{\tau}\phi
\nonumber\\
&+& \frac{\alpha \pi^2}{\beta^2}\ \int_0^\beta \int_0^\beta \rd\tau
\rd\tau'\ \frac{
\sin^2(\frac{\phi(\tau)-\phi(\tau')}{2})}{\sin^2(\frac{\pi}{\beta}(\tau-\tau'))}\, .
\label{action}
\eea
%
The first two terms  
account for the charging effect,
whereas the third, ``dissipative''  term describes the
electron tunneling.
Introducing a two component unit vector ${\boldsymbol S}=(\cos\phi\,,\,
\sin\phi)$ at  any point of the Matsubara circle, 
${\mathscr   A}[\phi]$ can be viewed as the classical energy of the XY model with
long range interaction [\onlinecite{Kosterlitz}].
The first term in Eq.\,\eqref{action}
is then identical to the spin wave approximation
to an extra short range interaction. Finally,  the 
term $\propto  n_G$ acts like a purely imaginary
external torque on the XY spins.

We are primarily interested in 
the ``magnetic'' susceptibility and effective charging energy of the model~\eqref{action}.
By magnetic susceptibility we mean the quantity
\bea\label{sksjsko}
\chi_m=\frac{2}{\beta^2}\ \int_{0}^{\beta}\int_{0}^{\beta}\rd\tau\rd \tau'\  \big\langle S_x(\tau)
S_x(\tau') \big\rangle\  ,
\eea
characterizing the linear response of the spin chain to an
external
``magnetic'' field   coupled to  $S_x\equiv\cos\phi$.
The effective charging energy, in turn, is defined as
\bea\label{salsalksalk}
E_C^*=-\frac{1}{2\beta}\ \frac{\partial^2}{\partial n_G^2}\,
\log(Z)\ .
\eea
In the  terminology of  QFT a   quantity like \eqref{salsalksalk}
is usually  referred to as  a {\it topological} susceptibility because of the
topological nature of the second term in Eq.\,\eqref{action}, and
the parameter 
\bea\label{lslala}
\theta=2\pi n_G
\eea
is called the topological angle.

When  both  $\beta E_C,\, \alpha\to \infty $  the model
\eqref{action}  
develops  a universal scaling behavior
characterized by the   energy scale  $E^*$ (Kondo temperature).
In this regime $g_0=(2\pi^2\alpha)^{-1}$ 
plays the role  of the bare perturbative coupling and 
the first term in Eq.\,\eqref{action} just provides an
explicit ultraviolet  cut-off of the dissipative action, with the cut-off energy 
\bea\label{lsaks}
\Lambda=2\pi^2\, E_C\, \alpha\ .
\eea
The Kondo temperature
admits  a  perturbative  expansion of the form [\onlinecite{Kosterlitz,Hofstetter}]
\bea\label{llsalka}
E^*
\simeq 2\pi^2E_C\ \alpha^2\,  
\re^{-\alpha\pi^2}\ \Big(1+\sum_{k=1}^{\infty}b_k\ \alpha^{-k}\
\Big)\ ,
\eea
where the symbol $\simeq$ stands for asymptotic equality.
While the overall factor $2\pi^2$ in Eq.\,\eqref{llsalka} is a matter of convention,
all the expansion  coefficients $b_k$ are determined unambiguously 
within the standard perturbation theory. 
In particular~[\onlinecite{Lukyanov}]
\bea\label{lskasalk}
b_1=-{ \frac{3}{ 8}}\ .
\eea

In the scaling limit, i.e., in the limit when $\Lambda\to\infty$, $\alpha\to\infty$ with
the energy scale $E^*$ kept fixed,
the  magnetic 
susceptibility~\eqref{sksjsko}  
is expected to possess the normal 
scaling behavior of the form [\onlinecite{Lukyanov}]
\bea\label{slslsalks}
\chi_m={ \frac{1}{2\pi^2\alpha}}\ \Big[\, F_m(\kappa,\theta)+
O\big(( \beta\Lambda)^{-1}\big)\, \Big]\ ,
\eea
where the  parameter $\kappa$  is the inverse temperature measured in
units of the Kondo temperature,
\bea\label{lssaalu}
\kappa=\beta E^*\ ,
\eea
and the topological angle $\theta$ is given by Eq.\,\eqref{lslala}.

The scaling behavior of the effective charging energy~\eqref{salsalksalk} 
is a  more
delicate issue. Let us note at this point that
field theories possessing topologically nontrivial classical Euclidean
solutions   (instantons)  usually suffer  from  specific ``instanton
divergences''. 
The phenomenon  was originally observed in  
4D QCD~[\onlinecite{Hooft}]. Later it was extensively studied in the
context of the 2D  
$O(3)$ nonlinear $\sigma$-model~[\onlinecite{Frolov,Berg,Luscher,Blatter}].
For $E_C=\infty$, the  model  \eqref{action} 
admits instanton solutions~[\onlinecite{Korshunov}]
which result in the ``small instanton problem'' (see 
the appendix below).
For large but finite $E_C$ the Coulomb 
term not only makes the theory perturbatively well-defined,
but also regularizes the ``small instanton divergence''. 
As $E_C\to\infty$,
the divergence restores  and entails
the anomalous scaling behavior for the effective charging 
energy  [\onlinecite{Lukyanov}]:
\bea\label{lsaklsk}
E_C^*=2\pi^2E^*\, \big[\, L\, \cos(\theta)
+F_t(\kappa,\theta)+o(1)\, \big] .
\eea
Here $L$ is a temperature-independent 
constant,   diverging  when  $\alpha\to \infty$  and $o(1)$ is a correction which vanishes in this limit. 
The scaling function $F_t$ is  defined modulo a transformation
$F_t\to F_t+c\ \cos\theta$, since any finite constant $c$ can be absorbed 
by the divergent term.
This  ambiguity can be resolved by imposing
a proper   normalization condition, say,
\bea\label{slasalsa}
\lim_{\kappa\to\infty}F_t(\kappa,0)=0\ .
\eea
As soon as  the normalization of $F_t$  is chosen, the
divergent term in Eq.~\eqref{lsaklsk} is  also defined unambiguously. 
According to the result of Ref.\,[\onlinecite{Lukyanov}]
\bea\label{slsals}
L(\alpha)=2\pi^2\alpha-5\, \log\alpha-C\ .
\eea
The constant $C$ has been found recently in Ref.~[\onlinecite{Luk}]:
\bea\label{slslsaj}
C=5\, \gamma_E+6\,\log 2+ 10\,\log\pi=18.492\ldots\ ,
\eea
where $\gamma_E=0.5772\ldots$ is Euler's constant.

As was argued in Refs.\,[\onlinecite{Lukyanov,Luk,LTZ}],
the scaling behavior of   thermodynamic functions  of the model~\eqref{action}
can be described  in terms of solutions of the
differential equation
\bea\label{Osrt}
\Big[\, -\partial_y^2+
\kappa^2\,\exp\big(\,\re^{y}\,\big)
-\kappa^2\sin^2(\theta)
\, \Big]\, \Psi(y) = 0\, .
\eea
Namely,
let  $\Psi_+$ and $\Psi_-$ be  solutions of Eq.\,\eqref{Osrt}
which are fixed by the asymptotic conditions
\bea\label{llisa}
\Psi_-(y)\to
  \re^{y\kappa\cos\theta}\, ,\ \  \ \ y\to-\infty
\eea
and
\bea\label{llauisa}
\Psi_+(y)\to 
\sqrt{\frac {\pi}{\kappa}}\, \exp\Big[
-{\textstyle \frac{1}{4}} \re^y-\kappa\,{\rm Ei}\big(\textstyle{\frac{\re^y}{2}}\big)\, \Big]
 ,\    \ y\to+\infty\, \eea
with ${\rm Ei}(z)=\dashint_{-z}^\infty\frac{\rd x}{x}\, \re^{-x}$.
Then the scaling function $F_t$ defined in Eq.\,\eqref{lsaklsk},
satisfying the normalization condition \eqref{slasalsa},
reads as follows~[\onlinecite{Luk}]:
\bea\label{lssa}
F_t(\kappa,\theta)=-\kappa^{-1}\ \partial_{\theta}^2\log {\bar Z}(\kappa,\theta)\ ,
\eea  
where
\bea\label{dsgfasta}
{\bar Z}(\kappa,\theta)&=&
\frac{ (2\re^{\gamma_E}\kappa^2)^{\kappa\cos\theta}}
{\Gamma(1+2\kappa\cos\theta)}\ 
  W[\Psi_+,\Psi_-]\  ,
\eea
and $W[\Psi_+,\Psi_-]=\Psi_+\partial_y\Psi_--\Psi_-\partial_y\Psi_+$ is
the Wronskian of the solutions $\Psi_+(y)$ and $\Psi_-(y)$.
It should be emphasized that
$\Psi_-$ can be defined by means of  the
asymptotic condition\ \eqref{llisa}
for    $0\leq \theta<\frac{\pi}{ 2}$ only.
For $\frac{\pi}{2}< \theta\leq \pi$,
the solution $\Psi_-$   grows at large negative $y$ and
the asymptotic formula \eqref{llisa}  does not define $\Psi_-$
unambiguously. It is possible to show that
the function
$\Psi_-/\Gamma(1+2\kappa\cos\theta)$  is an entire function
of the complex variable $\zeta=\cos\theta$ for  real  $y$. So
the solution $\Psi_-(y)$    can be
introduced within 
$\frac{\pi}{2}\leq \theta\leq \pi$
through analytic continuation with respect  to the variable $\theta$  from the
domain $0\leq \theta<\frac{\pi}{ 2}$.

The scaling functions $F_m(\kappa,\theta)$ defined in Eq.\,\eqref{slslsalks} 
were proposed for the cases $\theta=0$ and $\theta=\pi$ in 
Refs.\,[\onlinecite{Lukyanov}] and [\onlinecite{LTZ}], 
respectively. These results can be summarized as follows.
Let us introduce  the function 
\bea\label{lssklsk}
M(\kappa,\theta)=A_+(y)+A_-(y)+\frac{\partial_y A_+(y)-
\partial_y A_-(y)}{s_+(y)-s_-(y)}\, ,
\eea
where $s_\pm(y)=\partial_y\log\Psi_\pm(y)$
and
\bea\label{lsksksl}
A_+(y)&=&\int_{y}^{\infty}\frac{\rd u}{\Psi^2_+(u)}
\int_u^{\infty}\rd v\, \re^{v}\, \Psi^2_+(v)\ ,\nonumber\\
A_-(y)&=&
\int^{y}_{-\infty}\frac{\rd u}{\Psi^2_-(u)}
\int^u_{-\infty}\rd v\, \re^{v}\, \Psi^2_-(v)
\ .
\eea
Notice  that the RHS in Eq.\,\eqref{lssklsk} does not actually depend
on the choice of $y$.
Then
\bea\label{salsalsl}
F_m(\kappa,\theta)=
M(\kappa,\theta)\,  \ \ \ {\rm with}\ \ \ \theta=0\ \ {\rm and}\ \
 \pi\ .
\eea
It is essential to point out that no arguments have been given that relation
\eqref{salsalsl} holds true for any   values of $\theta$
different  from $\theta=0$ and $\theta=\pi$.

Let us now turn to the expression for the effective 
charging energy at zero temperature and gate charge.
Eqs.\,(\ref{lsaklsk}-\ref{slsals}) imply that
\begin{equation}
\frac{E_C^*}{E_C}\,
\bigg|_{n_G=\beta^{-1}=0}
\ \simeq\  f(\alpha)\ \re^{-\alpha \pi^2}\ \ \ {\rm as}\ \ \alpha\to\infty\ ,
\label{general}
\end{equation}
where
\bea\label{lssalsak}
f(\alpha)=4\pi^4\alpha^2\,
\big[\,1-{\textstyle{\frac{3}{8\alpha}}}+O(\alpha^{-2})\, \big]\, 
\big[\,L(\alpha)+o(1)\,\big]\, ,
\eea
and $L(\alpha)$ is given by Eq.\,\eqref{slsals}.

There is a controversy in the
literature about the asymptotic behavior of
$E_C^*$ at  $n_G=\beta^{-1}=0$ in
the large-$\alpha$ limit.
All the predictions agree on the
exponential suppression of the effective charging energy,
but there are numerous conflicting results  for
the leading term of the pre-exponential 
factor $f(\alpha)$, with powers ranging from $\alpha^1$ to $\alpha^{6.5}$
[\onlinecite{Panyukov&Zaikin, Wang&Grabert, Hofstetter, Larkin, Koenig}].
Several attempts to resolve the issue
by numerical simulations
[\onlinecite{Hofstetter, Wang&Egger, Herrero}] have failed
because of the difficulty to reach the large-$\alpha$ regime where
the exponential factor in Eq.\,\eqref{general} dominates.
Using efficient algorithms, it is possible to
compute the effective charging energy at essentially
zero temperature for $\alpha\lesssim 2$ [\onlinecite{Werner05}].
In this range of $\alpha$'s, there was no obvious agreement
with any of the theoretically predicted asymptotic formulas.
While the pre-exponential factors of Refs.\,[\onlinecite{Panyukov&Zaikin, Hofstetter, Koenig}]
could be ruled out on the basis of the numerical results
presented in Ref.~[\onlinecite{Werner05}], no definite conclusion
could be reached concerning the predictions of
Refs.\,[\onlinecite{Wang&Grabert, Lukyanov}]. It should be noted 
that   the  value of 
the  constant $C$ defined in Eq.\,\eqref{slslsaj} was not available at that time.

The goal of the present work is to  verify the  analytical results quoted  above,
including expression~\eqref{lssalsak}, 
by means of Monte Carlo simulations.

\section{Numerical Method}

In order to simulate the electron box, we map
the system (\ref{action}) 
for $n_G=0$ to a finite  chain of 
classical XY spins, 
by discretizing imaginary time into $N$ equal slices 
$\Delta\tau=\beta/N$,
\bea\label{discreteaction}
{\mathscr   A}_{\rm XY}[\phi]&=&\frac{1}{ 2E_C\Delta\tau}\ 
\sum_{k=1}^{N}\big(1-\cos(\phi_{k+1}-\phi_k)\, \big)\nonumber\\ 
&+&
\frac{\alpha \pi^2}{N^2}\ \sum_{k<k'}\frac{1-
\cos(\phi_k-\phi_{k'})}{\sin^2(\frac{\pi}{N}(k-k'))}\ .
\eea
The quasiperiodic boundary condition $\phi_{N+1}=\phi_1+2\pi w$  is   employed. 
In fact, the simulation treats the phases as 
compact variables and the winding number $w$ 
is computed by summing up the phase 
differences (between neighboring spins) along the chain. These phase 
differences are defined modulo $2\pi$ and we use 
the smaller angle -- a procedure which 
should yield accurate results for 
sufficiently small time steps (see also comments in section \ref{charging}).

The effective charging energy  at gate charge zero  
can be computed from the winding numbers
$w$ of the paths $\phi$  and the corresponding
weights $P_w$ -- the probabilities for a given $w$ [\onlinecite{Hofstetter}]:
\begin{equation}
\frac{E_C^*}{E_C}\, \bigg|_{n_G=0}=\frac{2\pi^2}{\beta E_C}\ \frac{\sum_w  w^2\, P_w}
{\sum_w P_w }\ .
\end{equation}
Results for non-zero gate charge can be obtained from the expectation values
$\langle\, \cdots\,  \rangle_w$ in the different winding sectors $w$:
\begin{equation}
\langle\, {\cal O}\, \rangle = \frac{\sum_w \re^{2\pi\ri w n_G }\ P_w\,
\langle\, {\cal O}\, \rangle_w}{\sum_w \re^{2\pi\ri w n_G}\ P_w }.\label{O_ng}
\end{equation}
Because of cancellation effects due to the phase factors in Eq.\,\eqref{O_ng}, 
the error bars can grow rapidly as a function of $n_G$, 
especially at small $\alpha$, where the 
$P_w$-distribution is very broad. 
This manifestation of a sign problem restricts the 
parameter range for which accurate results can be obtained.

We use Wolff cluster updates with efficient
treatment of long range interactions [\onlinecite{Wolff, Luijten&Bloete}]
in our Monte Carlo simulations.
This algorithm builds clusters in a time $O(N \log N)$ and allows us to simulate chains of up to $10^7$ spins.
For a discretization step $\Delta\tau E_C = 0.05$
it is therefore possible to reach temperatures down to $\beta E_C=5\cdot 10^5$.

\section{Results}

\subsection{Magnetic susceptibility}

\begin{figure}[t]
\centering
\includegraphics [angle=-90, width= 8.5cm] {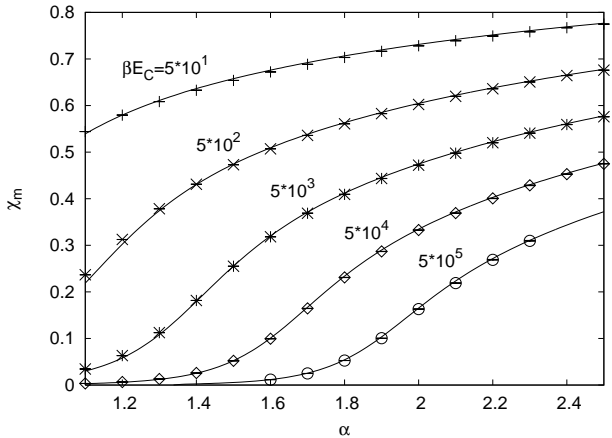}
\caption{ Comparison of $\frac{1}{  2\pi^2\alpha}\ M(\kappa, 0)$
defined in Eq.\,\eqref{lssklsk} (lines)
and Monte Carlo results (symbols) for $\chi_m$
(Eq.\,\eqref{sksjsko}, $n_G=0$).
The numerical data, which include error bars, were obtained for $\Delta\tau E_C=0.05$.
}
\label{m2_alpha}
\end{figure}

In Fig.\,\ref{m2_alpha} we compare the Monte Carlo results for the 
magnetic susceptibility for $n_G=0$ against 
the scaling form \eqref{slslsalks},\,\eqref{salsalsl}. 
The  plot for $n_G=\frac{1}{2}$ $(\theta=\pi)$
looks pretty much the
same as Fig.\,\ref{m2_alpha}.
Note that we  applied  relations
\eqref{llsalka},\,\eqref{lskasalk},\,\eqref{lssaalu} to find the
value of the scaling parameter $\kappa$
corresponding to a given $\alpha$ and $\beta E_C$.
The agreement between the analytical prediction and Monte Carlo results 
is within  $ \sim  0.5\%$  for 
$\beta E_C=5\times 10^n,\ n=1,\, 2,\, 3$ and within    $\sim 1-2\%$ for
$\beta E_C=5\times 10^n,\ n=4,\, 5$.
Among other things,
this indicates  that the asymptotic series \eqref{llsalka} truncated at $k=1$
approximates  the Kondo temperature $E^*$ with an estimated error  $\lesssim 1\%$ 
for any  $\alpha>1$. 
This level of accuracy seems reasonable given that the $k=1$ term leads to a correction of approximately $10\%$ for these $\alpha$.

As has been    mentioned 
above there is no strong theoretical argument  that 
relation \eqref{salsalsl} is  satisfied for  $\theta\in[0,\pi]$.
However, as shown in the appendix, 
the saddle point  approximation for  $\chi_m$ is 
consistent with the hypothesis that the relation 
holds true for any $\theta\in[0,\pi]$.
Further evidence that 
Eq.~\eqref{salsalsl} is satisfied for all $\theta$ comes from the study of the zero-temperature 
limit of the scaling function $F_m(\kappa,\theta)$. Namely,
as follows from\ \eqref{sksjsko},\,\eqref{slslsalks}, there exists a limit
\bea\label{lasksal}
\lim_{\kappa\to\infty}\big(\kappa\,F_m(\kappa,\theta)\big)=f_m(\theta)\ .
\eea
Here $f_m(\theta)$ is some function of 
the single variable $\theta$ which possesses
a singular behavior as $\theta\to \pi$. 
Using results of  Refs.~[\onlinecite{LTZ,Luk}] one can  show
that
\bea\label{lassjjh}
f_m(\theta)\to {\pi\over \pi-\theta}\ \ \ \ {\rm as}\ \ \ \pi-\theta\to+0\ .
\eea
At the same time, with  the WKB approximation  
it is not difficult to  prove the relation
\bea\label{slsalksls}
\lim_{\kappa\to\infty}\big(\kappa\,M(\kappa,\theta)\big)=
{\theta\over\sin(\theta)}\ .
\eea
Apparently Eqs.~\eqref{lasksal},\,\eqref{lassjjh} and
\eqref{slsalksls} 
are consistent with the 
conjecture $F_m(\kappa,\theta)=M(\kappa,\theta)$.

\begin{figure}[t]
\centering
\includegraphics [angle=-90, width= 8.5cm] {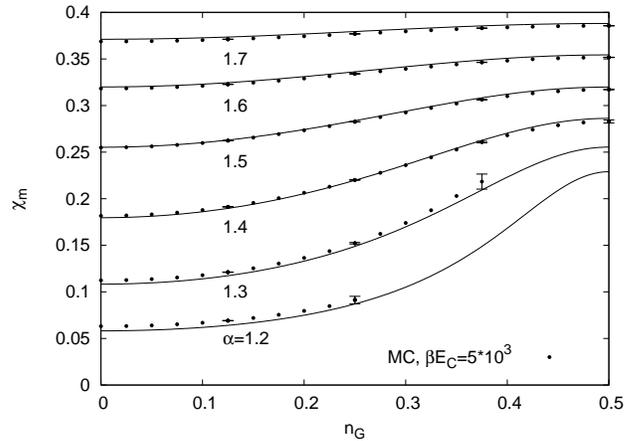}
\caption{$\frac{1}{  2\pi^2\alpha}\ M(\kappa, 2\pi n_G)$
defined in Eqs.\,\eqref{lssklsk}
(lines) and 
Monte Carlo results (symbols) for $\chi_m$ \eqref{sksjsko}
and several values of $\alpha$. 
The numerical data were obtained for $\beta E_C = 5\cdot 10^3$ 
and $\Delta\tau E_C=0.05$; error bars at $n_G=0.125$, 
0.25, 0.375 and 0.5 indicate the accuracy.}
\label{m2_ng}
\end{figure}

For the above
reasons, we found it natural
to test Eq.\,\eqref{salsalsl}   numerically. The result 
is  depicted in Fig.\,\ref{m2_ng}. 
The missing data points for 
$\alpha=1.2$ and 1.3 
fell victim to the sign problem mentioned in the previous section. This sign problem prevents us from investigating the singularity at $\theta=\pi$, which appears in the limit $T\rightarrow 0$.

\subsection{Effective charging energy for zero gate charge} 
\label{charging}

First of all it is instructive  to determine  a range   of temperatures
appropriate for numerical tests of the scaling relation~\eqref{lsaklsk}.
The leading
temperature-dependent correction to Eq.\,\eqref{lsaklsk}  readily follows from
the  saddle point calculations presented in the appendix.
Those calculations suggest (see formula \eqref{alsaaslk}) to
replace the temperature-independent  constant $L=L(\alpha)$~\eqref{slsals}  by
\bea\label{kjssks}
L(\alpha,\beta E_C)=L(\alpha)+L_1\big({\textstyle\frac{2\pi^2}{\beta E_C}}\big)\ ,
\eea
with
\bea\label{ssslsjssj}
L_1(x)=2\,\re^{\frac{x}{2}}\ {\rm Ei}(-x)+2\, \log\big(x\re^{\gamma_E}\big)\ .
\eea
For $\beta E_C=5\cdot 10^1$,
the term  $L_1$ turns out to be fairly large $(\approx 1.0)$.
For this reason, it makes no sense to use
Monte Carlo data  with $\beta E_C=5\cdot 10^1$
to test   the relation~\eqref{lsaklsk}.
In the case  $\beta E_C=5\cdot 10^2$, $L_1\approx 0.2$, so the correction
is small but still visible and it is useful   to take it into account.
For $\beta E_C=5\cdot 10^n$ with $n\geq 3$,
the term $L_1$ in Eq.\,\eqref{kjssks} is  
smaller than the error bars of our Monte Carlo data $(< 0.03)$,
and thus negligible.

In Ref.\,[\onlinecite{Werner05}] it was noted that
the disagreement between previous
Monte Carlo results [\onlinecite{Hofstetter, Wang&Egger, Herrero}]  for $E_C^*$ at
$\alpha\gtrsim 1$
were due to lattice effects.
Therefore it is very important to take proper care of the systematic
errors introduced by the imaginary time discretization in the calculation of $E_C^*$.

\begin{figure}[t]
\centering
\includegraphics [angle=-90, width= 8.5cm] {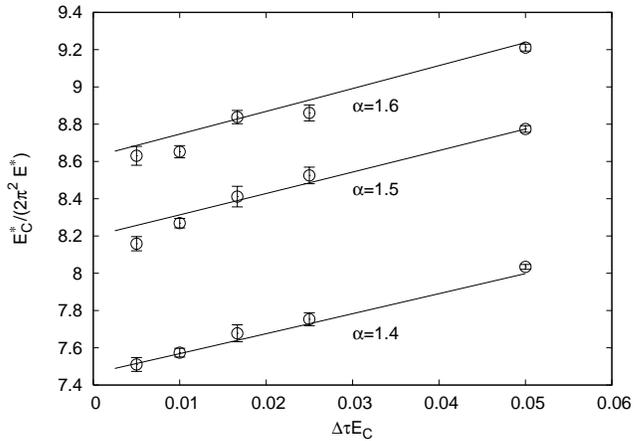}
\caption{Illustration of lattice effects for
$\frac{E_C^*}{2\pi^2E^*}\ (n_G=0)$.
The Monte Carlo data correspond
to $\alpha=1.4,\,1.5,\,1.6$ and  $\beta E_C = 5\cdot 10^3$.
The lines  show the analytical results corresponding to
the lattice fitting parameter ${\bar C}=2.8$ in Eq.~\eqref{kusdytsjssks}.}
\label{delta_tau}
\end{figure}

As a matter of fact, the strong effect of the  discretization on
the topological susceptibility is well known in the context of the
2D  $O(3)$ nonlinear $\sigma$-model [\onlinecite{Berg,Luscher,Blatter}]. 
The primary reason of this effect is that strictly speaking
the concept of winding number in a 
lattice formulation breaks down. In a discretized theory every
field configuration can be continuously transformed into any other. If
lattice configurations are sufficiently smooth, an unambiguous topological
charge may be assigned. Conversely, for field configurations containing large 
fluctuations the interpolation is not unique and 
the winding number definition becomes ambiguous.
For finite $E_C$
the discretized Coulomb term in the lattice action \eqref{discreteaction} 
suppresses the lattice
configurations containing large
fluctuations. However,
it is possible to show  that if 
$\alpha\,\Delta\tau E_C>1$
then starting with non-zero charged configurations
one can continuously lower the  lattice
action ${\mathscr   A}_{\rm XY}$  
to zero by a local minimization in the spin variables~[\onlinecite{LUT}].
For this  reason  the divergent constant $L$ in  Eq.\,\eqref{lsaklsk}
(which is not universal and  depends
on details of the  discretization of the action~\eqref{action})
is a   singular  function of the lattice parameter
\bea\label{slsskh}
\delta=\alpha\,\Delta\tau E_C\ ,
\eea
at $\delta= 1$. Namely,
\bea\label{lsjlsa}
L\propto -2\, \log(1-\delta)\ .
\eea 
In Fig.\,\ref{delta_tau},  $\frac{E_C^*}{2\pi^2E^*}$ with $n_G=0$ is plotted
as 
a function of $\Delta\tau E_C$ for $\beta E_C = 5\cdot 10^3$ and
some values of $\alpha$.
An analysis of the  plotted  Monte Carla data   suggests that the
effect of the lattice discretization for $\delta\ll 1$ can be accounted for by means of the 
following modification of Eq.\,\eqref{kjssks}: 
\bea\label{kusdytsjssks}
L(\alpha,\beta E_C, \delta)&=&
L(\alpha)+L_1\big({\textstyle\frac{2\pi^2}{\beta E_C}}\big) \nonumber \\ &&
-2\, \log(1-\delta)+2\, {\bar C}\,\delta\ .
\eea
Here $L(\alpha)$ and $L_1(x)$  are given by Eqs.\,\eqref{slsals} and \eqref{ssslsjssj},
respectively, and
${\bar C}$ is a fitting parameter. 
The continuous lines 
in  Fig.\,\ref{delta_tau} correspond to  ${\bar C}=2.8$.

\begin{figure}[t]
\centering
\includegraphics [angle=-90, width= 8.5cm] {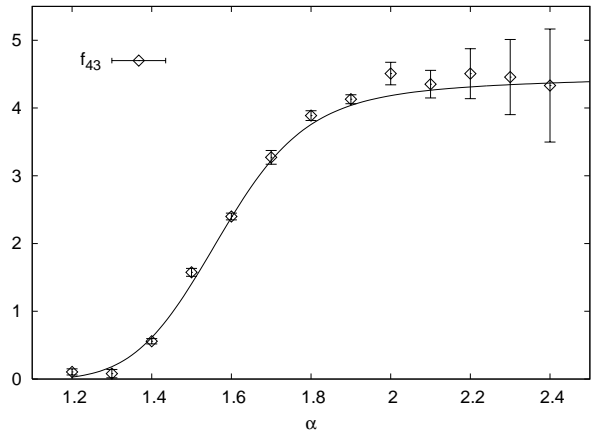}
\caption{Comparison of 
the Monte Carlo data and analytical 
prediction for $f_{43}$ defined in Eq.\,\eqref{def_f_nm}. 
The numerical data correspond to 
$\Delta\tau E_C=0.05$ and the analytical 
result was obtained using Eqs.\,\eqref{sjksa},\,\eqref{lssa}.}
\label{f_43}
\end{figure}

In order to minimize the impact of
the non-universal temperature and
lattice effects discussed above,
we first study the quantity $(n_G=0)$
\begin{equation}
f_{nm}(\alpha) = \left.\frac{E_C^*}{2\pi^2 E^*}\right|_{\beta E_C=5\cdot 10^n}-
\left.\frac{E_C^*}{2\pi^2 E^*}\right|_{\beta E_C=5\cdot 10^m}\ .
\label{def_f_nm}
\end{equation}
As follows from Eq.\,\eqref{lsaklsk}, the anomalous terms
cancel out for such differences.  Hence
one should   expect  that the functions  $f_{nm}$  possess the normal scaling
behavior of the form
\bea\label{sjksa}
f_{nm}(\alpha)\to F_t(\kappa_n,0)-F_t(\kappa_m,0)\ \
(\, n,\,m, \,\alpha\to\infty)
\eea
with  $\kappa_{n,m}$
the values of the scaling parameter $\kappa$ \eqref{lssaalu}
corresponding to
a given $\alpha$ and  $\beta E_C=5\cdot 10^{n,m}$.
Also notice that
the non-universal temperature corrections to the scaling behavior \eqref{sjksa}
are  expected to be sufficiently  small provided $n,m\geq 3$.
Since the most accurate Monte Carlo data are
those for $n=3$, we plot in Fig.\,\ref{f_43} the result
for $f_{43}$, obtained for a discretization step $\Delta\tau E_C=0.05$, and
the analytical prediction (Eqs.\,\eqref{sjksa},\,\eqref{lssa}).
The agreement is as good as it can be expected, given the
errors on the Monte Carlo data, and a similar result is found for $f_{32}$.

\begin{figure}[t]
\centering
\includegraphics [angle=-90, width= 8.5cm] {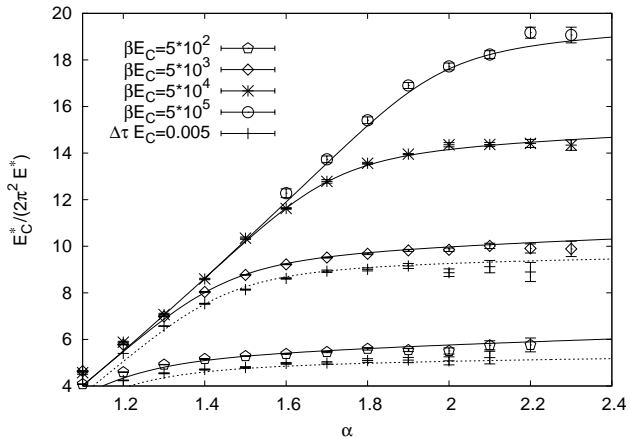}
\caption{Analytical 
prediction (lines) and Monte 
Carlo results (symbols) for $\frac{E_C^*}{2\pi^2E^*}$ 
and several values of $\beta E_C$ ($n_G=0$). 
The numerical data were 
obtained for $\Delta\tau E_C=0.05$ 
(points on solid lines) and $\Delta\tau E_C=0.005$ (points on dotted lines). 
The coefficient ${\bar C}=2.8$ \eqref{kusdytsjssks} was used for all the analytical curves.}
\label{E_C_alpha}
\end{figure}

In Fig.\,\ref{E_C_alpha} the prediction~\eqref{lsaklsk},\,\eqref{lssa} 
for $n_G=0$ and  $L$  understood as
in Eq.\,\eqref{kusdytsjssks}, is compared to the Monte Carlo data for different
values of the parameters  
$\beta E_C$ and $\Delta\tau E_C$.  
The value $\bar C=2.8$ has been used 
for all analytical curves. 
The agreement between the theoretical 
predictions and the Monte Carlo results 
shows that the scaling formulas \eqref{lsaklsk},\,\eqref{slsals},\,\eqref{lssa} 
for the effective charging energy are correct. 
In particular, this implies that 
the leading power in the pre-exponential 
factor $f(\alpha)$ (see Eqs.\,(\ref{general}) and (\ref{lssalsak})) is $\alpha^3$.

\section{Summary}
 
We have presented 
scaling formulas for the single electron box 
in the limit of almost perfect transmission, 
as well as a conjecture for the gate charge dependence of 
the magnetic susceptibility. These analytical predictions have been 
tested by means of Monte Carlo simulations of the 
discretized action (corresponding to an XY spin chain with long range couplings) 
with up to $10^7$ spins. 
Because of the lattice effects, it was necessary to introduce a 
constant,
$\bar C=2.8$,  in order 
to compare the analytical predictions to the simulation results for
the effective charging energy. 
But this single fitting parameter was 
enough to obtain a good agreement between the Monte Carlo data and the 
theoretical predictions over a whole range of temperatures 
($5\cdot 10^2 \le \beta E_C \le 5\cdot 10^5$), 
lattice spacings ($0.005 \le \Delta\tau E_C \le 0.05$) and 
dissipation strengths ($1.2 \lesssim \alpha\lesssim 2.4$). 
These results essentially verify the scaling formulas for the effective charging energy for $n_G=0$
presented in Ref.\,[\onlinecite{Luk}],
and thus also settle the controversy
surrounding the pre-exponential factor $f(\alpha)$ in Eq.\,\eqref{general}.
Unfortunately the quality of the Monte Carlo data for $E_C^*(n_G)$ with $n_G\ne 0$ was not sufficient for a meaningful test of the $\theta$-dependence of the scaling function $F_t(\kappa,\theta)$  \eqref{lsaklsk} against the theoretical prediction \eqref{lssa}.                                                                         

As for the magnetic susceptibility,
our investigation   suggests that some of the 
results  from  Refs.\,[\onlinecite{Lukyanov,LTZ}]
for $n_G=0$ and $n_G=1/2$  can be generalized to all values of the gate charge.

\section*{Acknowledgments}

The calculations have been performed on 
the Hreidar Beowulf cluster at ETH Z\"urich, using the ALPS library 
[\onlinecite{ALPS}]. 
The authors thank M. Troyer for the generous allocation of computer time.
SL is grateful to Alexander B. Zamolodchikov for previous collaborations and his interest in this
work.

\noindent
Research of SL is supported in part by DOE grant
$\#$DE-FG02-96 ER 40949.
PW acknowledges support from NSF DMR 0431350.

\appendix

\section{\ Saddle point approximation}

In this appendix we consider  
the saddle point  approximation
for the model \eqref{action} in the limit  $E_C\to\infty$.
In the absence of the first term,  the local   minima 
of the Euclidean  action ${\mathscr   A}[\phi]$  in the sectors
with winding number $w\geq 0$ are saturated  by
the instanton solutions [\onlinecite{Korshunov, Nazarov}]
\bea\label{lslsjshd}
\re^{\ri\phi_{\rm inst}(\tau)}=\re^{\ri \phi_0}\ 
\prod_{m=1}^{w}\frac{z-z_m}{1-z_m^*z}\ \ \ {\rm  with}\ \ \ 
z=\re^{2\ri\pi\tau/\beta}\, .
\eea
The set of complex parameters $\{z_m\, :\,|z_m|<1\}_{m=1}^w$ and
the real  constant  $0\leq\phi_0<2\pi$ 
are  instanton moduli.
Let us introduce  an external magnetic field in the system:
\bea\label{lslkslk}
{\mathscr   A}_h[\phi]={\mathscr   A}[\phi]+h\sqrt{\alpha}\ 
\int_0^\beta\rd\tau\, \cos\big(\phi (\tau)\big)\ .
\eea
We assume that $h$ is sufficiently small so  that
the additional term  has no appreciable
effect on the instanton solutions. 
Calculating the action ${\mathscr   A}_h[\phi]$ for the paths \eqref{lslsjshd} yields
\bea\label{slissal}
{\mathscr   A}_h[\phi_{\rm inst} ]&=&\ri\,\theta w+ \alpha \pi^2\ w+
{\textstyle\frac{1}{ 2}}\, \beta h\sqrt{\alpha}\ (-1)^{w}\times\nonumber\\
&& \Big(\,\re^{\ri\phi_0}
\prod_{m=1}^w z_m+\re^{-\ri\phi_0}
\prod_{m=1}^w z^*_m\, \Big)\ ,
\eea
where  $\theta$ is the topological angle \eqref{lslala}.
Within the saddle point  approximation the
partition function   is given by
\bea\label{slasasa}
Z_{\rm inst}(h)=\sum_{w=0}^{\infty}
\int_{w} \rd{\cal M}_w\ 
\re^{-{\mathscr   A}_h[\phi_{\rm inst} ]}\ ,
\eea
where   $\int_{w} $ stands for the integral over 
the domain 
$0\leq\phi_0<2\pi,\ |z_m|<1\ (m=1,\ldots w)$.
The integration   measure $\rd{\cal M}_w$
should be computed as usual by
integrating out the Gaussian fluctuations
around the instanton solutions. It was calculated 
in Refs.\,[\onlinecite{Larkin,Feigelman}]:
\bea\label{sklal}
\rd{\cal M}_w&=&
\frac{(\beta \Lambda)^w}{ w!}\ (2\pi^2\alpha)^{\frac{1}{2}+w}\times 
\\ &&\frac{\rd\phi_0}{2\pi}\wedge\prod_{m=1}^m
\frac{\rd z_m\wedge\rd z_m^*}{2\pi\ri }
\ \ {\rm det}\parallel G(z_i,z_j)\parallel\,  .\nonumber
\eea
Here
\bea\label{slalsk}
G(z_i,z_j)=\frac{1}{1-z_iz_j^*}\ \,,
\eea
and $\Lambda$ is the perturbative ultraviolet  cut-off \eqref{lsaks}.

Expression \eqref{slasasa} can be evaluated in closed form. 
Indeed, let us expand the exponent $\re^{-{\mathscr   A}_h[\phi_{\rm inst} ]}$
into a power series of $\beta h$ and then integrate explicitly over   $\phi_0$.
This yields
\bea\label{slslksa}
Z_{\rm inst}(h)=\sqrt{2\pi^2\alpha}\, 
\sum_{p=0}^{\infty}\frac{1}{ (p!)^2} \Big(\frac{\beta h\sqrt{\alpha}}{ 2}
\Big)^{2p}\, 
Z_p\big(\kappa\re^{-\ri\theta}\big) ,
\eea
with $\kappa=\beta\, \Lambda\alpha\re^{-\alpha\pi^2}$
and
\bea\label{ytsksks}
Z_p(\lambda)&=&1+\sum_{w=1}^{\infty}\frac{\lambda^w}{ w!}\ 
\prod_{m=1}^w\int_{|z_m|<1}\times\\
&&\prod_{m=1}^w
\frac{\rd z_m\wedge\rd z_m^*}{2\pi\ri }\ (z_mz_m^*)^p
\  \ {\rm det}\parallel G(z_i,z^*_j) \parallel .\nonumber
\eea
It is easy to see that $Z_p(\lambda)$ can be interpreted
as  a  (normalized) fermionic partition function:
\bea\label{lskjssl}
Z_p(\lambda)&=&
\frac{\int{\cal D}{\bar c}{\cal D} c\ \re^{ - ({\bar c}\, {\hat G}^{-1}\, c)-
\lambda ({\bar c}, c)}}{
\int{\cal D}{\bar c}{\cal D} c\ \re^{- ({\bar c}\, {\hat G}^{-1}\, c)}}=\\ 
&&\frac{{\rm det} ({\hat G}^{-1}+\lambda)}{
{\rm det} ({\hat G}^{-1})}= {\rm det}\big(1+\lambda\, {\hat G}\big)\  ,\nonumber
\eea
where the operator ${\hat G}$ acting on the Grassman fields
is defined by the equation
\bea\label{slsalskj}
&&({\hat G} c)(z,z^*)=\\ &&\ \ \ \ \ \ \ \  
\int_{|\zeta|<1}\frac{\rd \zeta
\wedge\rd  \zeta^*}{2\pi\ri }\ (\zeta \zeta^*)^p\  G(z,\zeta)\, c(\zeta,
\zeta^*)\  .\nonumber
\eea
Hence,
\bea
\label{laslaoi}
\log Z_p= {\rm Sp}\log(1+\lambda {\hat G})=-
\sum_{s=1}^{\infty}{\textstyle\frac{(-1)^{s}}{ s}}\ {\rm Sp}({\hat G}^s),
\eea
with
\bea\label{sosduislks}
&&{\rm Sp}({\hat G}^s)= \prod_{m=1}^s\bigg[
\int_{|z_m|<1}\frac{\rd z_m\wedge \rd { z}^*_m}{ 2\pi\ri}\ (z_mz_m^*)^p\bigg]\times
\nonumber\\
&& \ \ \ \ \ \ \big[(1-z_1z^*_2)(1-z_2z^*_3)\ldots
 (1-z_s z^*_1)\,\big]^{-1}\ .
\eea

The small instanton problem is explicit now.
Indeed,  equation~\eqref{sosduislks} for $s=1$ can be written in the form
\bea\label{lksjsl}
{\rm Sp}({\hat G})=\int'_{|z_1|<1}\frac{\rd z_1
\wedge \rd z^*_1}{2\pi\ri(1-z_1 z^*_1)}
-\gamma_E-\psi(1 + p),
\eea
where $\psi(u)=\frac{\rd}{\rd u}\,\log\Gamma(u)$
and the   integral    diverges as $|z_1|\to 1$.
This limit corresponds to small instantons -- when $|z_1|$ is
close to one, the   solution $\phi_{\rm inst}$ \eqref{lslsjshd} with  $w=1$
is almost constant everywhere in  the segment $[0,\beta]$
except for a small neighborhood of the point
$\tau=\frac{\beta}{2\pi}\,{\rm arg}(z_1)$ where the instanton is localized.
Any regularization (denoted by $\int'$ in \eqref{lksjsl}) 
is essentially an instruction to suppress
contributions from the small-size instantons. 
As a matter of fact,
the Coulomb term  in the action \eqref{action} 
provides such a suppression  for any finite  $\beta E_C\gg 1$.
Thus
the     integral in \eqref{lksjsl}
should be understood as  [\onlinecite{Larkin}]:
\bea\label{asjsj}
\int'\ldots=\int_{|z_1|<1}\frac{\rd z_1
\wedge \rd  z^*_1}{ 2\pi\ri}\
\frac{\re^{-{\mathscr   A}^{(1)}_C}}{ 1-z_1 z^*_1}\ .
\eea
Here
\bea\label{jsshs}
{\mathscr   A}^{(1)}_C=\frac{\pi^2}{\beta E_C}\ \frac{1+z_1z^*_1}{
1-z_1z_1^*}
\eea
is the value of the Coulomb term
calculated on
the one instanton solution $(w=1)$. Then
\bea\label{alsaaslk}
{\rm Sp}({\hat G})={\textstyle\frac{1}{2}}\, \big[ L+
L_1\big({\textstyle\frac{2\pi^2}{\beta E_C}}\big) \big]+
\log(\kappa)-\psi(1 + p),
\eea
where $L=2\pi^2\alpha+O(\log\alpha)$ is some divergent
temperature -independent constant. 
The function $L_1(x)$ is given by Eq.\,\eqref{ssslsjssj}.
It vanishes in the limit  $\beta E_C\to \infty$.

All  integrals \eqref{sosduislks} with  $s>1$  are finite.
Integrating over the phases of $z_m$  in \eqref{sosduislks}  first,
one obtains  $(x_m=|z_m|^2$),
\bea\label{kjshakh}
{\rm Sp}({\hat G}^s)&=&\bigg[\prod_{m=1}^s\int_{0}^1\rd x_m \, x^p_m\bigg]\ 
\frac{1}{ 1-x_1\cdot\ldots\cdot x_k}=\nonumber\\
&&\sum_{n=0}^{\infty}
\bigg[\int_{0}^1\rd x\ x^{n+p}\bigg]^s=\zeta_p(s)\ ,
\eea
where $\zeta_p(s)=
\sum_{n=0}^{\infty}(n+p)^{-s}$ is the generalized Riemann $\zeta$-function.
Finally, $Z_{\rm inst}(h)$
can be  expressed in terms of the conventional
Bessel function,
\bea\label{slsalsl}
Z_{\rm inst}(h)=\re^{\frac{1}{2} L\lambda}\ \ \sqrt{2\pi^2\alpha}\ 
\Big({\frac{2\kappa}{ \beta h\sqrt{\alpha}}}
\,\Big)^{\lambda}
\, I_{\lambda}(\beta h\sqrt{\alpha}),
\eea
with $\lambda=\kappa\re^{-\ri\theta}$.

Some explanations   are in order at this point.
First, the corrections to the  saddle point formula \eqref{slasasa}
are small in $\alpha^{-1}$ but
they contain  ultraviolet divergent
integrals.
Taking these divergences properly into account leads
to a  renormalization of the bare
coupling $g_0=(2\pi^2 \alpha)^{-1}$
in Eq.~\eqref{slsalsl}. The renormalization
trades  $g_0$
for the
``running coupling constant''
$g(\kappa)$ subject to the RG flow equation [\onlinecite{Kosterlitz,Hofstetter}],
\bea\label{slssk}
\kappa\, \frac{\rd  g}{\rd \kappa}=2\, g^2+4\, g^3+O(g^4)
\ .
\eea

Second, in the above   derivation   we took into account
the instantons  with positive winding numbers only.
One can  indeed  ignore
the ``anti-instanton'' configurations corresponding to
$w<0$  provided $\Im m(\theta)\gg 1$.
Obviously, the contribution of anti-instanton fluctuations
dominates when  $\Im m(-\theta)\gg 1$. 
In this case the saddle point approximation leads to
\eqref{slsalsl} with
$\lambda=\kappa\re^{\ri\theta}$.
Both limiting  cases are neatly captured
by the same  formula  if   $\lambda$
is understood as
\bea\label{qewsaklsak}
\lambda=2\kappa\ \cos\theta\ .
\eea
Equation \eqref{qewsaklsak}  
is not easy to justify on  general grounds for $\theta\sim 1$, but
can be supported by means of  the exact result in the case $h=0$.

According to Ref.\,[\onlinecite{Luk}], the scaling behavior of
the  partition function~\eqref{partitionfunction} looks as follows
\bea\label{lkspaai}
Z\to Z_{\rm scal}=\re^{L\kappa\cos\theta}\ {\bar Z}(\kappa,\theta)\ \ \ 
(\beta E_C,\ \alpha\,\to\infty)
,
\eea
where $L$ and ${\bar Z}$ are given by~Eqs.\,\eqref{slsals} and~\eqref{dsgfasta}, respectively.
We consider now the high-temperature (small-$\kappa$) expansion of \eqref{lkspaai}. 
Here it is useful to  introduce
the running coupling constant $g=g(\kappa)$ as a solution of the
equation
\bea\label{slssal}
\kappa=g^{-1}\ \re^{-\frac{1}{2g}}\ .
\eea
Note that $g\ll 1$ and it
solves
the RG flow equation \eqref{slssk} within the two-loop approximation.
After a change of variables
\bea\label{kssajk}
y=g\, x-\log g\ ,
\eea
the differential equation~\eqref{Osrt} can be written in the form
\bea\label{siuslsks}
\Big[ 
-\partial_x^2+\re^{x}+(g\kappa\cos\theta)^2+\delta U(x) \, \Big]\, \Psi(x) = 0\, ,
\eea
where 
\bea\label{kljsjy}
\delta U(x)=\exp\Big(\,\frac{\re^{g x}-1}{g}\,\Big)-\re^{x}-(g\kappa)^2\ .
\eea
For $|x|\sim 1$ the term $\delta U$ is $O(g)$,
so the solution $\Psi_+$ is approximated by the Macdonald function
\bea\label{ajksaijhns}
\Psi_+(x)= 
2\,g^{\frac{1}{2}}\  K_{2g\kappa \cos\theta}(2\re^{\frac{x}{2}})\ \big(\,1+O(g)\,\big)\ .
\eea
The normalization factor in front of $K$ is fixed by
matching \eqref{ajksaijhns} to the asymptotic condition~\eqref{llauisa}.
As $(-x)\gg 1$,  $\Psi_+$ takes the form
\bea\label{lkjldl}
\Psi_+(x)\to C_+\ \re^{x\kappa g \cos\theta}+C_-\ \re^{-x\kappa g \cos\theta}\ .
\eea
The coefficient $C_-$ is simply related to the Wronskian in Eq.\,\eqref{dsgfasta},
namely
\bea\label{lasjsjl}
W[\Psi_+,\Psi_-]=2\kappa\cos\theta\, g^{-\kappa\cos\theta}\ C_-\ .
\eea
In view of \eqref{siuslsks}, one can show 
that the high-temperature expansion of   $Z_{\rm scal}$~\eqref{lkspaai} has the form:
\bea\label{lsjlsiuys}
Z_{\rm scal}&\simeq& \re^{\kappa (L-\log g)\cos\theta}\ \ 
\frac{\kappa^{2\kappa\cos\theta}}{\sqrt{g}}\times \nonumber\\ &&
\sum_{m=0}^{\infty} g^m\ Z_m(\kappa,\theta)\  ,
\eea
where the  coefficients  
$Z_m(\kappa,\theta)$ admit systematic expansions in powers of $\kappa$.
In particular
\bea\label{lsjlss}
Z_{0}(\kappa,\theta)=
\frac{(2 \re^{\gamma_E})^{\kappa\cos\theta}}
{\Gamma(1+2\kappa\cos\theta)}\ .
\eea
It is straightforward to check  that
the saddle point partition function $Z_{\rm inst}(0)$
is in agreement with \eqref{lsjlss} provided
$\lambda$ in equation~\eqref{slsalsl} is given by \eqref{qewsaklsak}
and the effect of the renormalization of  $g_0=(2\pi^2\alpha)^{-1}$ is
taken into account.

Let us return to  Eqs.\,\eqref{slsalsl},\,\eqref{qewsaklsak}
to find
the magnetic susceptibility within the saddle point approximation:
\bea\label{slsa}
\chi_m^{(\rm inst)}&=&
\frac{2}{ \beta^2}\
\partial^2_h\log Z_{\rm inst}(h)\big|_{h=0}=\\ &&
\frac{1}{2\pi^2\alpha}\ \, \frac{1}
{ g(\kappa)\ (1+ 2\kappa\cos\theta)}\, .\nonumber
\eea
Again the effect of renormalization has been  accounted for in  \eqref{slsa}.
It is easy to see that
formula \eqref{slsa} is in agreement with
the hypothesis
that relation~\eqref{salsalsl} is satisfied for any $\theta$.
Indeed,   as follows from the above analysis,
in the regime  $\kappa\ll 1$ the potential $U(y)=\kappa^2\,\exp\big(\,\re^{y}\,\big)
-\kappa^2\sin^2(\theta)$
in the  differential equation \eqref{Osrt} has the effect of a rigid wall at some
point $y_0\approx -\log g$, i.e., to the left from
this point, the potential is negligible, but to the
right, it grows very fast, so the solution $\Psi_+(y)$ is
essentially zero. More precisely, when $y$ is below but close
to $y_0$ the solution $\Psi_+$ is approximated by a linear function,
$\Psi_+(y)\approx g^{-\frac{1}{2}}\, (y_0-y)$.
Within this ``rigid wall'' approximation
$A_+(y_0)\approx 0$,
$\partial_y A_+(y_0)\approx 0$ and $s(y_0)\gg1 $.
Therefore, the RHS of Eq.\,\eqref{lssklsk}, calculated at $y=y_0$,
can be approximated  by  $A_-(y_0)$.
In the calculation of  $A_-(y)$,
the solution  $\Psi_-$ needed for the integral \eqref{lsksksl} can be
replaced by its asymptotics \eqref{llisa}.
This yields the relation 
\bea\label{lskslk}
M(\kappa,\theta) \approx 2\pi^2\alpha\ \chi_m^{(\rm inst)}\ ,
\eea
where $\chi_m^{(\rm inst)}$ is given by \eqref{slsa}.


\begin{thebibliography}{99}
\bibitem{LUT}
A very similar  phenomenon was originally observed by
M. L\"uscher [\onlinecite{Luscher}]
in the $O(3)$ nonlinear $\sigma$-model.
\end{thebibliography}
\end{document}